\documentclass[aps,prl,twocolumn]{revtex4}
\usepackage{amsmath,bm,epsfig,,subfigure}
\usepackage{color}
\usepackage[normalem]{ulem}

\def\Fbox#1{\vskip1ex\hbox to 8.5cm{\hfil\fboxsep0.3cm\fbox{%
  \parbox{8.0cm}{#1}}\hfil}\vskip1ex\noindent}  

\newcommand{\C}[1]{{\mathcal{#1}}}    
\let \= \equiv \let\*\cdot \let\~\widetilde \let\-\overline

\let \= \equiv \let\*\cdot \let\~\widetilde \let\^\widehat \let\-\overline

\begin{document}
\title{Derivation of the Johnson-Samwer $T^{(2/3)}$ Temperature Dependence of the Yield Strain in Metallic Glasses}
\author{Ratul Dasgupta, Ashwin Joy, H.G.E.Hentschel$^*$ and Itamar Procaccia}
\affiliation{Dept. of Chemical Physics, The Weizmann Institute of
Science, Rehovot 76100, Israel
\\$^*$Dept of Physics, Emory University, Atlanta, Georgia, }

\date{\today}
\begin{abstract}
Metallic Glasses are prone to fail mechanically via a shear-banding instability. In a remarkable
paper Johnson and Samwer demonstrated that this failure enjoys a high degree of universality in the sense that a large group of metallic glasses appears to possess a yield-strain that decreases with
temperature following a $-T^{2/3}$ law up to logarithmic corrections. In this Letter we offer a theoretical derivation of this law. We show that our formula fits very well simulational data on typical amorphous solids.
\end{abstract}
\maketitle
{\bf Introduction}: A satisfactory derivation of the rheology of amorphous solids like metallic glasses under various mechanical and magnetic external strains is still far from being accomplished. Examples of universal phenomena and universal relations are rare and far between \cite{12DKP}. One of these rare
examples is the Johnson-Samwer $T^{2/3}$ law \cite{05JS} which pertains to the temperature dependence of the yield-strain (the value of the strain where the material fails via plastic instabilities).
Examining a large group of metallic glasses these authors proposed the law
\begin{equation}
\gamma_{_{\rm Y}}(T,\dot\gamma) = \gamma_{_{\rm Y}}(T=0,\dot \gamma=0) \left\{1-\left[AT\ln(\omega_0/C\dot\gamma)\right]^{2/3}\right\} \ , \label{JS}
\end{equation}
where $\gamma_{_{\rm Y}}(T=0,\dot \gamma=0)$ is the yield-stress at athermal quasi-static conditions,
$\dot\gamma$ is the external strain rate, $A$ and $C$ are material constants and $\omega_0$ is a
microscopic inverse time scale. Johnson and Samwer offered a derivation of this law \cite{05JS} based on Frenkel's theory \cite{26Fre} which is appropriate for an infinite crystals having a periodic elastic
energy density, leaving a derivation which is proper for an amorphous solid for the future. This task
was picked up in Ref. \cite{10CCL}. These authors recognized that in amorphous solids plastic yielding
follows a saddle-node bifurcation where an eigenvalue $\lambda_P$ of the system's Hessian hits zero at
a value of the strain $\gamma=\gamma_P$.
Such a bifurcation leads to a local rearrangement which is characterized by an energy barrier $\delta E$ which scales like $\delta E\sim (\gamma_P-\gamma)^{3/2}$. Assuming that this energy barrier is much larger than $kT$ ($k$ being Boltzmann's constant), Ref. \cite{10CCL} showed that a temperature induced
barrier crossing would result in a law of the form of Eq. \ref{JS}. In this Letter we argue that the energy barrier that needs to be crossed to achieve shear localization is sub-extensive; on the other hand the energy barrier associated with a single localized plastic instability is minute, and in the thermodynamic limit it scales like $1/N$ where $N$ is the number of particles \cite{10KLP}. Therefore the conditions at which the system yields via a coherent shear localized band need to be reconsidered. 
\begin{figure}
\includegraphics[scale = 0.15]{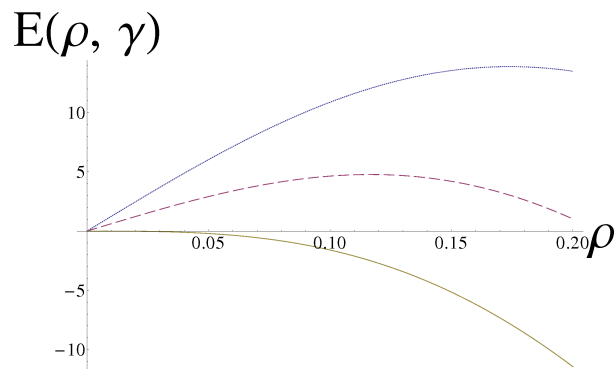}
\caption{(Color Online). The total plastic energy Eq. (\ref{Erg}) for the creation of an array of quadrupoles with density
$\rho$ for three values of $\gamma$: $\gamma=\gamma_{_{\rm Y}}-0.1$ (upper curve),
$\gamma=\gamma_{_{\rm Y}}-0.05$ (middle curve),
and $\gamma=\gamma_{_{\rm Y}}$ (lower curve). In the present case $\gamma_{_{\rm Y}}= 0.07$. To generate this picture we use the measured constants $\cal E\approx$~37.2, $\nu\approx 0.31$, $\epsilon^*\approx 0.082$ and $a=1.83$. Finally $U_p\approx 0.22$.}
\label{Eplast}
\end{figure}

{\bf Yielding via shear localization}: In recent work it was argued that when the external strain
approaches the shear-localization yield-strain the nature of plastic instabilities can change qualitatively \cite{12DHPa,12DHPb}. At low external strains a plastic instability results in a local rearrangement such that the non-affine displacement of particles can be very well modeled by the displacement field associated with a {\em single} Eshelby inclusion in an elastic matrix \cite{57Esh}. In 2D this field assumes a quadrupolar structure.
At larger values of the strain the instability results in 2D in a highly correlated array of Eshelby
quadrupoles that are aligned at 45 degrees to the principal stress axis. All the quadrupoles are
in phase and in total they result in shear localization in a narrow strip. One could show that this highly correlated array is a minimum energy state which depends on the density $\rho$ of the quadrupoles. For $\gamma<\gamma_{_{\rm Y}}$ the only solution is $\rho=0$, i.e. isolated qudrupoles,
but at $\gamma=\gamma_{_{\rm Y}}$ a bifurcation opens up a new solution with a finite density, cf. Fig. \ref{Eplast}.
Under conditions of athermal quasistatic straining (AQS), it was shown that the yield strain $\gamma_{_{\rm Y}}$ is given by the expression \cite{12DHPb}
\begin{equation}
\label{gammay}
\gamma_{_{Y}} \equiv \gamma_{_{Y}} (T=0,\dot \gamma = 0) \equiv \frac{\epsilon^*}{2(1-\nu)}.
\end{equation}
where $\epsilon^*$ is the eigenstrain induced by single plastic instability in the background elastic matrix whose Poisson's ratio is $\nu$. Each quadrupole can be very well modeled as an Eshelby inclusion with eigenstrain $\epsilon^*$ and core size $a$ \cite{57Esh}. The energy density cost of creating a linear array of $\C N$ quadrupoles all with the same orientation, separated by distance $R=L/{\C N}$ (in a 2-dimensional system of size $L^2$) was computed analytically \cite{12DHPb} in the form
\begin{equation}
\frac{E(\rho,\gamma)}{La} = U_p\bigg[\left(1-\frac{\gamma}{\gamma_{_{Y}}}\right)a\rho - B(a\rho)^3 + C(a\rho)^5\bigg]
\label{Erg}
\end{equation}
where $U_p = [\C E\pi (\epsilon^{*})^2]/[4(1-\nu^2)]$ with $\C E$ being Young's modulus; while $B= 4\zeta(2)$ and $C =6\zeta(4)$, where the Riemann zeta functions are respectively $\zeta (2) = \pi^2/6$ and $\zeta (4) = \pi^4/90$. This energy is shown in Fig. \ref{Eplast}. Examining this energy
we realize that in AQS conditions yield via shear localization can occur only at $\gamma=\gamma_{_{\rm Y}}$. With finite temperature we can have thermally-assisted transitions which we consider next.

{\bf Thermally assisted plastic yield}: Considering the yield stress at any finite temperature $T\ne 0$; we realize that thermal fluctuations can always (if given enough time) cross any energy barrier and therefore under conditions of quasistatic straining the yield stress should always vanish
\begin{equation}
\label{gammayt}
\gamma_{_{Y}} (T, \dot \gamma =0) =0.
\end{equation}
Accordingly, for strictly quasistatic straining $\dot \gamma =0$ the limit $T\rightarrow 0$ yield-stress is different from the $T=0$ yield-stress. This jump in the limit was observed in \cite{05JS} and was
interpreted as a ``quantum effect". There is no need to invoke quantum mechanics to understand this
simple issue.

\begin{figure}
\includegraphics[scale = 0.20]{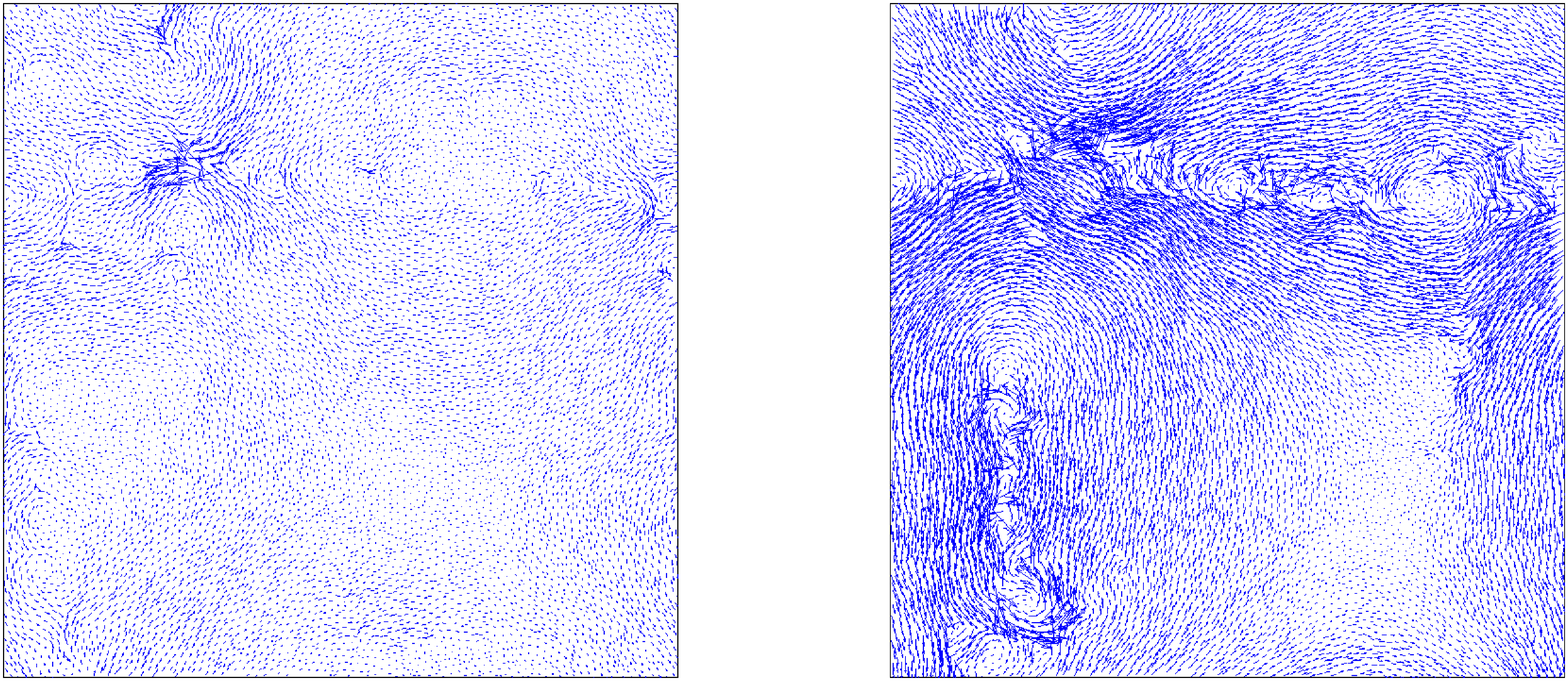}
\caption{(Color Online). Visualization of the process of shear localization in the AQS simulations
whose full details can be found in Ref. \cite{12DHPb}. Even with the quasistatic protocol with arbitrary long waiting times if necessary one cannot resolve the formation of the full structure of shear localized band of quadrupoles (right panel) from the creation of a single quadrupole (left panel). The right panel appears instantly after the left panel.}
\label{instant}
\end{figure}

In reality one always strains at some finite value of $\dot \gamma$. Then thermal fluctuations can lead to barrier crossing if the timescale for such crossing $\tau$ is smaller or similar to the straining time scale $\dot \gamma^{-1}$, leading to the condition for yielding due to thermal fluctuations given by
\begin{equation}
\label{times}
\tau \dot \gamma = O(1).
\end{equation}
If we had a process in which every particle in the system acted independently, it would be enough to compute the energy barrier per particles, say $U_b$, and compare it to the thermal energy, to be used  in the Arrhenius form
\begin{equation}
\label{arrhenius}
\tau = \tau_0 \exp{[U_b/kT]},~ \text {for independent particle motion}
\end{equation}
where $\tau_0$ is the inverse of the attempt frequency for barrier crossing.
In fact, the situation here is more complex. The way that shear localization occurs in practice is that there exists a mode that gets localized on $N^*\ll N$ particles which are involved in the first quadrupole that forms and then the rest of the linear array of qudrupoles forms instantly, as seen in Fig. \ref{instant}. Thus the rate determining step is the creation of the first quadrupole, and we estimate the appropriate times scale as
\begin{equation}
\label{arrhenius2}
\tau = \tau_0 \exp{[N^*U_b/kT]}, ~ \text {for concerted particle motion}
\end{equation}
Combing these results gives the expression for yielding as
\begin{equation}
\label{yielding}
\tau_0 \dot \gamma \exp{[N^*U_b/kT]} = O(1).
\end{equation}
Note that $N^*$ remains independent of $N$ in the thermodynamic limit, representing the number of
particles involved in the concerted barrier crossing of the creation of one quadrupolar non-affine
displacement field. We expect $N^*$ to be of the order of 100, give or take a factor of 2.

We now use these result to estimate how finite temperatures and finite strain rates change the value of the AQS yield strain due to shear localization.

{\bf Calculating the energy barrier}:
Examining Eq. (\ref{Erg}) we see that the plastic energy density due to shear localization at strain $\gamma$ can be written in terms of the dimensionless variable $x=\rho a$. The height of the barrier {\em per particle} $U_b$ is determined by the energy density $E^{*}/La$ at $x^{*} = \rho^{*}a$ which can be found from $\partial (E/La)/\partial x =0$ with the result that the barrier occurs at
\begin{equation}
\label{x}
x^{*} = \rho^{*}a = \sqrt{[1-\gamma/\gamma_{_{Y}}]/2}/\pi,
\end{equation}
and its strip energy density is given by
\begin{equation}
\label{ee}
E^{*}/La = \frac{\sqrt{2} U_p}{3\pi} [1-\gamma/\gamma_{_{Y}}]^{(3/2)}.
\end{equation}
Eq.~(\ref{ee}) implies that the barrier energy per particle in the strip is
\begin{equation}
\label{u}
U_b = (E^{*}/La)/n = [\sqrt{2} U_p/(3\pi n)] [1-\gamma/\gamma_{_{Y}}]^{(3/2)},
\end{equation}
where $n$ is the number density of particles in the strip of dimension $La$.

 Finally we need to estimate $\tau_0$, the inverse attempt frequency. There are two candidates. The first is the eigenvalue of the Hessian matrix $\lambda_p$ which is associated with the eigenfunction that gets localized. The square root of $\lambda_p$ is an inverse time scale.
 This time scale however diverges to infinity when $\gamma$ is close to $\gamma_{_{\rm Y}}$, and the purely thermal time scale becomes shorter and more relevant.  We thus argue here that the relevant
  time scale can be estimated from the thermal fluctuations of the individual particles in the solid
 which determines the time scale of a concerted motion of $N^*$ particles. For a single particle the  thermal time scale is $l/v$ where $l\approx 1/\sqrt{n}$ is the typical distance between particles and the typical velocity $v$ can be found from the equipartition theorem $m \langle v^2\rangle = kT$. Thus for $N^*$ particles moving together we estimate
\begin{equation}
\label{tau0}
\tau_0 = N^*\sqrt{m/(n kT)}.
\end{equation}
Now combining Eq.~(\ref{yielding}) for the yield strain with Eq.~(\ref{u}) for the barrier height and Eq.~\ref{tau0} for the bare time scale we finally find
\begin{equation}
\label{yieldtg}
\gamma_{_{Y}} (T, \dot \gamma )/\gamma_{_{Y}} = 1 - \left[\frac{3\pi n kT}{\sqrt{2}N^* U_p}\right]^{2/3} \log^{2/3}\left[ \frac{\sqrt{n kT/m}}{N^*\dot\gamma} \right].
\end{equation}
We note that at this point only $N^*$ is not known with certainty, only in order of magnitude. The expression has the $T^{(2/3)}$ temperature reduction in the yield strain found experimentally by Johnson and Samwer, cf. Eq. (\ref{JS}), including the logarithmic correction term.

{\bf Comparison with numerical simulations}:
We have performed 2D Molecular Dynamics simulations on a binary system which is an excellent glass former and is known to have
a quasi-crystalline ground state. Each atom in the system is labeled as either ``small''(S) or ``large''(L) and all the particles interact via Lennard
Jones (LJ) potential. All distances are normalized by $\sigma_{SL}$, the distance at which the LJ potential between the two species becomes zero and
the energy is normalized by $\epsilon_{SL}$ which is the interaction energy between two species. For detailed information on the model potential and
its properties, we refer the reader to Ref \cite{87WSS}. The number of particles taken in all our simulations is 100489 at a number density $n = 0.985$
with a particle ratio $N_L/N_s = (1+\sqrt{5})/4$. The mode coupling temperature $T_{MCT}$ for this system is known to reside close to $0.325
\epsilon_{SL}/k_B$. All particles have identical mass $m_0$ and hence the time is normalized to $ t_0 = \sqrt{\epsilon_{SL}{\sigma_{SL}}^2/m_0}$.
For the sake of computational efficiency, the interaction potential is smoothly truncated to zero along with its first two derivatives at a cut-off
distance $r_c = 2.5 \sigma_{SL}$. To prepare the glasses, we start from a well equilibrated liquids at a high temperature of $1.2 \epsilon_{SL}/k_B$
which are supercooled to $0.35 \epsilon_{SL}/k_B$ at a reduced quenching rate of $3.4 \times 10^{-3} t_0^{-1}$. We then equilibrate these supercooled
liquids for times greater than $10\tau_{\rm rel}$, where $\tau_{\rm rel}$ is the  time taken for the self intermediate scattering function to become $1\%$ of its initial
value. Following this equilibration, we quench these supercooled liquids deep into the glassy regime at a temperature of $0.03\epsilon_{SL}/k_B$ at a
much slower quenching rate of $3.2 \times 10^{-6} t_0^{-1}$.
\par
\begin{figure}
\includegraphics[scale = 0.25]{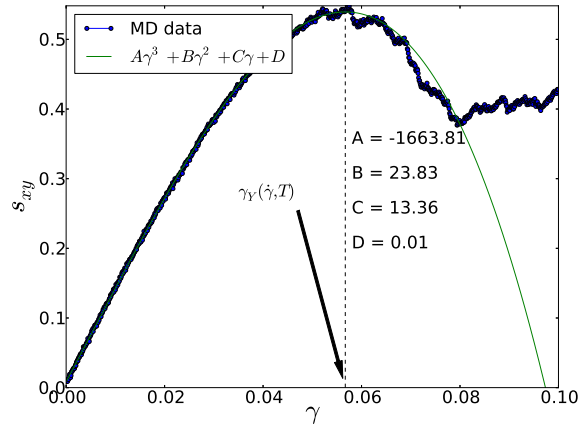}
\caption{(Color Online). Typical stress vs. strain curve obtained at finite temperature and
finite strain rate. The yield stress was estimated by fitting a cubic to the curve (see inset)
and finding the maximum of the curve.}
\label{sigvsgam}
\end{figure}

We perform simple shear loading experiments by integrating the SLLOD system of equations \cite {96CCC} along with Lees-Edwards periodic boundary
conditions \cite{72LE}. As can be expected, work done on the system will lead to heat dissipation and hence we keep our system connected to the thermostat in
order to keep the temperature constant throughout the mechanical deformation. We employ a strain rate $\dot{\gamma} = 10^{-5}$ in all our loading
experiments. A typical stress vs. strain curve is shown in Fig. \ref{sigvsgam}.
The data of stress vs strain were fitted to a cubic in $\gamma$ and the yield strain was estimated
from the maximum of the curve.

\begin{figure}
\includegraphics[scale = 0.25]{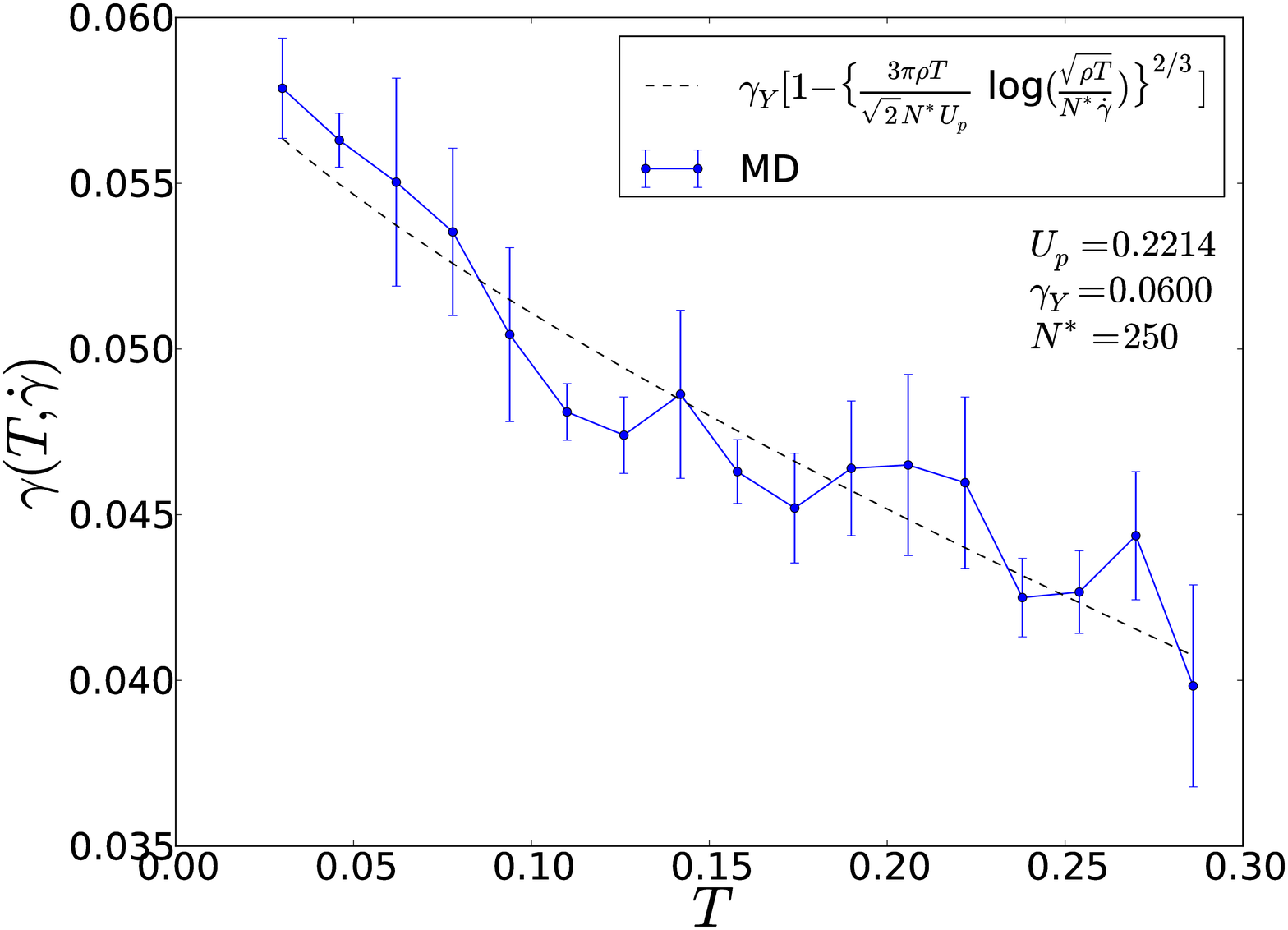}
\caption{(Color Online). Comparison of the simulational data of $\gamma(T,\dot\gamma)$ to the prediction Eq. (\ref{yieldtg}). The dotted line was obtained by finding the best fit for $N^*$. The error bars indicate averaging over three independent realizations.}
\label{fit}
\end{figure}

To compare with the theory above we need also $\gamma_Y$ at zero temperature and the value of $U_p$. We obtained these by minimizing the energy of one of the glasses generated as explained above to $T=0$.
Performing athermal quasi-static simulations on this sample we determined all the parameters appearing
in Eq.(\ref{yieldtg}), with the result $\gamma_Y\approx 0.06$ and $U_p=0.22$. Armed with these and all the other known numbers we compare the data for the temperature dependent yield strains to the
prediction of Eq. (\ref{yieldtg}) using $N^*$ to get a best fit, which is obtained for $N^*\approx 250$. The fit is shown in Fig. \ref{fit}.
Having in mind the inherent unknown factors hidden in the estimates Eqs. (\ref{times}),(\ref{arrhenius}) and (\ref{tau0}), we find the fit very satisfactory.

In summary, we have used the analytically computed energy associated with a correlated series
of quadrupolar structures that add up to a shear localizing instability to study the thermally
assisted yield strain associated with plastic failure in two dimensions. This energy is sub-extensive
and its strip-density remains unchanged in the thermodynamic limit. We could derive the Johnson-Samwer
$T^{2/3}$ law essentially without a free parameter except for an uncertainty regarding the value
of $N^*$. The fitted value of this number appears to be in the right order of magnitude, lending
strong support to the approach detailed above.

{\bf Acknowledgments}: This work had been supported in part by an ERC ``ideas" grant and by the Israel Science Foundation. Discussions with Eran Bouchbinder are gratefully acknowledged.

\end{document}